\documentclass[prb,twocolumn,showpacs]{revtex4}
\usepackage{times}
\usepackage{bm}
\usepackage[dvips]{graphicx}
\usepackage{amsmath}
\usepackage{bm}
\usepackage{natbib}
\usepackage{color}
\usepackage{epstopdf}
\begin{document}
%\preprint{}
\title{Insulating Behavior of an Amorphous  Graphene Membrane}
\author{Dinh Van Tuan,$^{1}$ Avishek Kumar,$^{2}$ 
Stephan  Roche,$^{1,3}$ Frank Ortmann,$^{1}$ M. F. Thorpe,$^{2}$ Pablo Ordejon$^{4}$}
\affiliation{
$^1$CIN2 (ICN-CSIC) and Universitat Aut\'{o}noma de Barcelona, Catalan Institute
 of Nanotechnology, Campus UAB, 08193 Bellaterra, Spain\\ 
$^2$Department of Physics and Astronomy, Arizona State University, Tempe, AZ 85287, USA\\
$^3$ICREA, Instituci\'{o} Catalana de Recerca i Estudis Avan\c{c}ats, 08070 Barcelona, Spain\\
$^4$Centre d'Investigaci\'o en Nanoci\`encia i Nanotecnologia - CIN2 (CSIC-ICN), Campus UAB, 
08193 Bellaterra, Spain}
\date{\today}
\begin{abstract}
We investigate the charge transport properties of planar amorphous graphene that is fully topologically disordered, in the form of  $sp^{2}$ three-fold coordinated networks consisting of hexagonal rings, but also including many pentagons and heptagons distributed in a random fashion. Using the Kubo transport methodology and the Lanczos method, the density of states, mean free paths and semiclassical conductivities of such amorphous graphene membranes are computed. Despite a large increase in the density of states close to the charge neutrality point, all electronic properties are dramatically degraded, evidencing an Anderson insulating state caused by topological disorder alone. These results are supported by Landauer-B\"uttiker conductance calculations, which show a localization length as short as 5 nanometers.
\end{abstract} 

\pacs{72.80.Vp, 73.63.-b, 73.22.Pr, 72.15.Lh, 61.48.Gh} 
\maketitle

%------------------------------------------------------------------------------

The physics of disordered graphene is at the heart of many fascinating properties such as Klein tunneling, weak antilocalization or anomalous quantum Hall effect (see reviews \cite{Review,ReviewSR}). The precise understanding of individual defects on electronic and transport properties of graphene is currently of great interest \cite{Krasheninnikov}. For instance, graphene samples obtained by large-scale production methods display a huge quantity of structural imperfections and defects that jeopardize the robustness of the otherwise exceptionally high charge mobilities of their pristine counterparts \cite{Novoselov2011}.  Indeed, the lattice mismatch-induced strain between graphene and the underlying substrate generates polycrystalline graphene with grain boundaries that strongly impact transport properties \cite{Ferreira}. However, despite the large amount of disorder, such graphene flakes (when deposited onto oxide substrates) usually maintain a finite conductivity down to very low temperatures owing to electron-hole puddles-induced percolation effects that preclude localization phenomena close to the Dirac point \cite{DassarmaRMP}. The predicted Anderson localization in two-dimensional disordered graphene has been hard to measure in non intentionally damaged graphene, in contrast to chemically modified graphene \cite{Moser,leconte}. Nevertheless, in a recent experiment it was possible to screen out electron-holes puddles using  sandwiched graphene in between two boron-nitride layers, together with an additional graphene control layer \cite{Ponomarenko2011}. As a result of puddles screening, a large increase of the resistivity was obtained at the Dirac point, evidencing an onset of the Anderson localization regime. 

Beyond individual defects and polycrystallinity, a higher level of disorder can be induced on graphene to the point of obtaining two-dimensional fully amorphous networks composed of $sp^2$ hybridized carbon atoms. Such networks contain rings other than hexagons in a disordered arrangement. The average ring size is six according to Euler's theorem, allowing these systems to exist as flat 2D structures. Experimentally, such amorphous two-dimensional lattices have been obtained in electron-beam irradiation experiments \cite{Kotakoski,meyer}, and directly visualized by high resolution electron transmission microscopy. Previously, indirect evidence for the formation of an amorphous network was obtained using Raman spectroscopy in samples subject to electron-beam irradiation \cite{Teweldebrhan}, ozone exposure \cite{Tao} and ion irradiation \cite{zhou}. In all these cases, an evolution from polycrystalline to amorphous structures was observed upon increase of the damage treatment.  In \cite{zhou}, further evidence of the formation of an amorphous network was obtained through transport measurements. These indicate the transition from a weak localization regime in the polycrystalline samples to variable range hopping transport in the strongly localized regime for amorphous samples, as evidenced by the temperature dependence of the conductivity. Localization lengths were estimated to be of the range 0.1 to 10 nm in the amorphous samples, depending on the degree of amorphization. From the theoretical side, models of the amorphous network have been proposed using stochastic quenching methods \cite{Kapko}, and molecular dynamics \cite{Holmstrom,Li,Lherbier2012}. Electronic structure calculations show that the amorphization yields a large increase of the density of states in the close vicinity of the charge neutrality point \cite{Kapko,Holmstrom,Li}. Despite the expected reduction of the conduction properties due to strong localization effects, Holmstr\"om {\em et al.} \cite{Holmstrom} suggest that disorder could enhance metallicity in amorphized samples, in contrast with the experimental evidence.

In this Letter, we explore the transport properties of two-dimensional $sp^{2}$ lattices with a massive amount of topological disorder, encoded in a geometrical mixture of hexagons  with pentagon and heptagon rings. The calculations are done using two complementary approaches: a Kubo formulation in which the conductivity of bulk 2D amorphous graphene lattices is determined, and a Landauer-B\"uttiker formulation where the conductance of stripes of amorphous graphene contacted to semi-infinite pristine graphene electrodes is calculated. Both approaches lead to similar findings. Depending on the ratio between odd versus even-membered rings, a transition form a graphene-like electronic structure to a totally amorphous and smooth electronic distribution of states is obtained.  The stronger the departure from the pristine graphene, the more insulating is the corresponding lattice, which transforms into a strong Anderson insulator with elastic mean free paths below one nanometer and localization lengths below 10 nm close to the charge neutrality point. Those structures are therefore inefficient to carry any sizable current, and unsuitable for practical electronic applications such as touch screens displays or conducting electrodes, but interesting for scrutinizing localization phenomena in low dimensional materials.

\begin{figure}[htbp]
\begin{center}
\leavevmode
\includegraphics[scale=0.3]{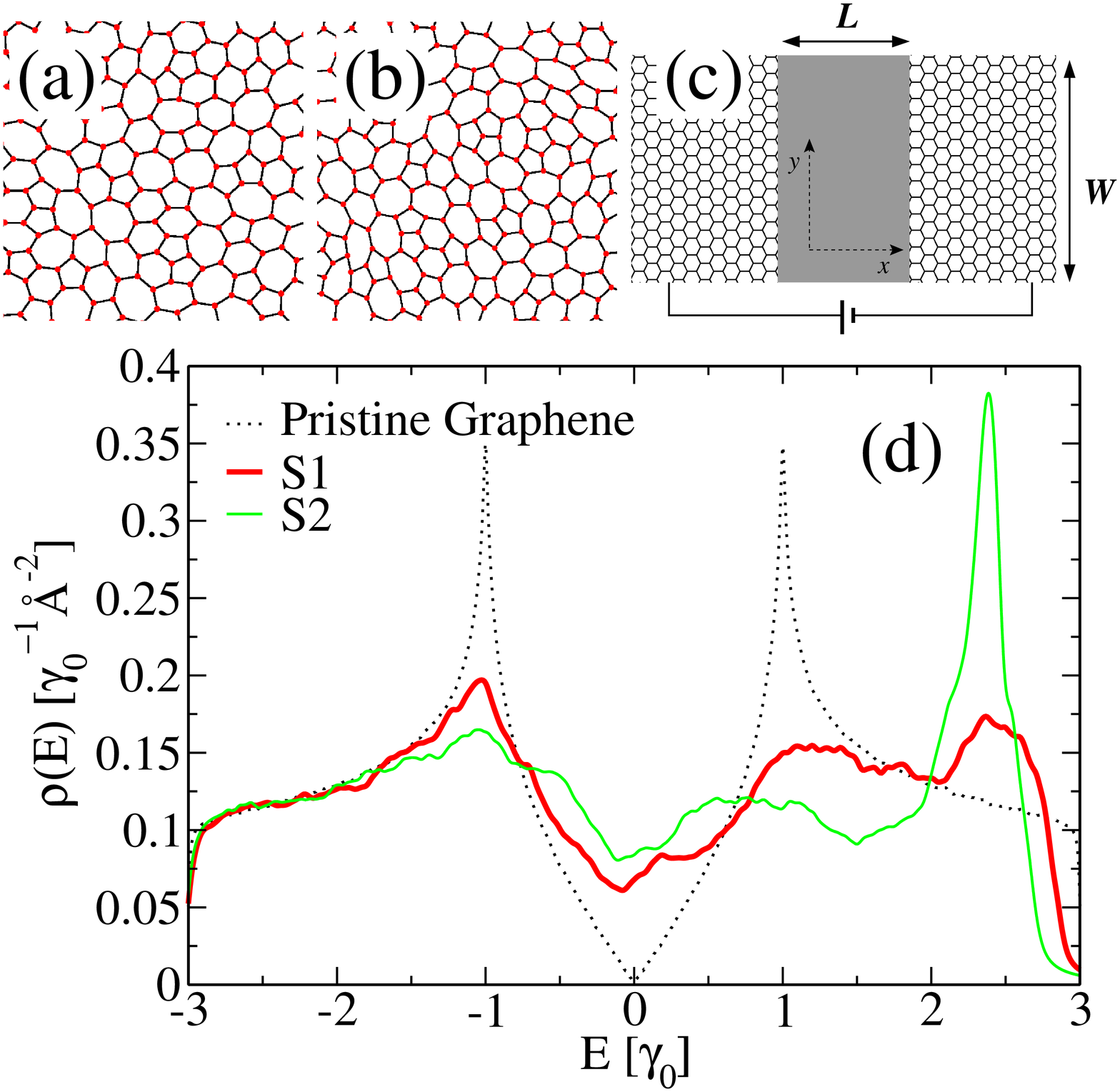}
\caption{(color online).  (a), (b) Details of  amorphous graphene samples S1 and S2, respectively, used to compute the conductivity with the Kubo approach. (b) Scheme of setup for the Landauer-B\"utikker calculations. The grey area represents the amorphous graphene stripe.
Periodic boundary conditions  are used in the $y$ direction. The graphene
electrodes are semi-infinite in the $x$ direction. (d) Total density of states of the two amorphous samples. The pristine crystalline graphene case (dashed lines) is also shown for comparison.}
\label{dos}
\end{center}
\end{figure}

\textit{Models of amorphous graphene}.- Amorphous models of graphene are prepared using the Wooten-Winer-Weaire (WWW) method \cite{6, 7}, introducing Stone-Wales defects \cite{8} into the perfect honeycomb lattice. These networks can be considered as the amorphous versions of the Haeckelite structures proposed by Terrones {\em et al.} \cite{terrones}. To generate the structures, periodic boundary conditions are imposed and the entire network is relaxed with the Keating-like potential \cite{9,Kapko}. 
A further relaxation of these models using forces from Density Functional Theory leads to slight changes in the bonding distances and angles, but to virtually identical radial distributions functions and, most importantly, electronic properties like the density of states, as shown in Ref.~\onlinecite{Li}. Fig. \ref{dos}(a)-(b) show pieces of the two studied sample S1 and S2, which respectively contain 10032 and 101640 atoms (all of them with three-fold coordination as the
honeycomb lattice, but topologically distinct). The amorphous character of the samples is demonstrated by analyzing the radial distribution function, as shown in Ref.~\onlinecite{Kapko}. Table \ref{specs} shows the parameters that characterize the two samples. For sample S1, 24$\%$ of the elementary rings are pentagons, 52$\%$ hexagons and 24$\%$ heptagons, while sample S2 has a larger share of odd-membered rings.
%: 44$\%$ of pentagons, 12$\%$ hexagons and 44$\%$ heptagons. 
In both samples, the number of heptagons is the same as that of pentagons, as required by Euler's theorem, and these systems can exist without an overall curvature as flat 2D structures with some distortions of bond lengths and angles, although they may pucker under some circumstances. We will only be concerned with the planar structures here.
%The root mean square angular deviation from the mean of $120\deg$ is  $11.02\deg$ for 
%sample S1 and $18.09\deg$ for sample S2. The second moment of the ring statistics 
%$< n^{2}>-<n>^{2}$ is $0.47$ for sample S1 and $0.88$ for sample S2. The Fermi energy 
%is $0.03\gamma_0$ for sample S1 and $0.05\gamma_0$ for sample S2. 
Sample S2 is an extreme case, having very few hexagons and being furthest from the pristine honeycomb lattice. This is useful to accentuate the differences between crystalline and amorphous samples, and to gain perspective. Nevertheless, it is likely that sample S1 is nearer to physical reality as it is less strained. 

\begin{table}    
\caption{Characteristic parameters for the two samples of amorphous graphene, S1 and S2} 
\label{specs}                                                          
\begin{tabular}{lcc}                                                             
\hline                 
\hline  
 &~~~~~~~~S1~~~~~~~~&~~~~~~~~S2~~~~~~~~\\
\hline
number of atoms & 10032 & 101640\\
\% of $n$-membered rings ($n=5/6/7$) & 24/52/24 & 44/12/44\\
$< n^{2}>-<n>^{2} $ & 0.47 & 0.88\\
RMS deviation of bond angles & 11.02$^\circ$ & 18.09$^\circ$ \\
RMS deviation of bond lengths & 0.044~\AA & 0.060~\AA\\
Fermi energy (in units of $\gamma_0$) & 0.03 & 0.05\\
\hline
\hline                                                                                   
\end{tabular}   
\label{table}                                                                             
\end{table}

For the calculation of the Landauer-B\"uttiker conductance, we set up models in which an amorphous stripe is contacted by two pristine graphene electrodes separated by a distance $L$, as shown in Fig. \ref{dos}(c).
We use models with increasing values of $L$, to explore the dependence of the
conductance on the length of the amorphous contact in the transport direction.
The models are periodic in the direction perpendicular to the  stripe, with a periodicity of $W$=11.4 nm, and they have the same ring statistics as the bulk sample S1 described above.

\begin{figure}[htbp]
\begin{center}
\leavevmode
\includegraphics[scale=1.0]{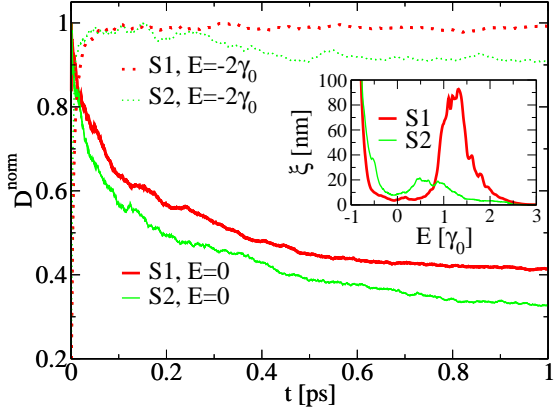}
\caption{(color online) Normalized time-dependent diffusion coefficients for two selected energies for both samples S1 and S2. Inset: localization lengths as a function of the carrier energy}
\label{difconst}
\end{center}
\end{figure}

%By varying the ring statistics, one will determine the scaling behavior of the transport properties with increased the level of amorphousness.

\textit{Electronic Properties}.- The electronic and transport properties of these disordered lattices are investigated using  $\pi$-$\pi$* orthogonal tight-binding (TB) model with nearest neighbor hopping $\gamma_0 = 2.8$ eV \cite{Kapko} and zero onsite energies. No variation of the hopping elements with disorder is included in the model as bond-length variation does not exceed a few percent ({\em cf.} Table \ref{table}); all dependence on disorder stems from the ring statistics, which is expected to be the dominant effect \cite{harrison}.  Fig. \ref{dos}(d) shows the density of states (DOS) of the two disordered samples, together with the pristine case (dashed line) for comparison. Sample 1, which keeps 52$\%$ of hexagonal rings, displays several noticeable features, similar to those found in previous studies \cite{Kapko,Holmstrom}. First, the DOS at the charge neutrality point (Fermi level) is found to be increased by a large amount~\cite{suplinf}. Additionally, the electron-hole symmetry of the band structure is broken by the presence of odd-membered rings which generate quasibound states at resonant energies \cite{Lherbier2011}. The hole part of the spectrum is still reminiscent of the graphene DOS, with a smoothened peak at the van Hove singularity, while for the electron part a second maximum appears close to the upper conduction band edge.  By reducing further the ratio of even versus odd-membered rings (sample S2), the second maximum develops to a strong  peak at about $E=2.5 \gamma_0$ while spectral weight at $E=3\gamma_0$ is suppressed. The redistribution of DOS at the upper conduction band edge is a signature of odd-membered rings and its strength with increasing number of such rings relates the statistical distribution of rings with the DOS features.

\textit{Transport Methodology}.- To explore quantum transport in these topologically disordered graphene samples,  a real-space order-$N$ quantum wavepacket evolution approach is employed to compute the Kubo-Greenwood  conductivity \cite{Kubo1966}.  Such method, pioneered in \cite{Roche1}, has been successfully applied to many different types of systems, and in
particular it has allowed to scrutinize Anderson localization
in oxygen functionalized graphene~\cite{leconte}.  The zero-frequency conductivity for carriers at energy $E$ is computed as
\begin{equation}
\sigma_{dc}=
%{
e^{2} 
%\over 2} 
\rho(E) {\displaystyle \lim_{t\to\infty}} \frac{d}{dt}\Delta X^{2}(E,t)
\label{eq:kubo}
\end{equation}
where $\rho(E)$ is the density of states and $\Delta X^{2}(E,t)$ is the mean quadratic displacement of the wave packet at energy $E$ and time $t$:
\begin{equation}
\Delta X^{2}(E,t) = \frac{\displaystyle {\large\rm Tr}\bigl[ \delta(E-{\cal H}) | \hat{X}(t)- \hat{X}(0) |^2 \bigr]}
{ \strut\displaystyle {\large\rm Tr}[\delta(E-{\cal H})]}
\label{DeltaX2}
\end{equation}
A key quantity in the analysis of the transport properties is the diffusion coefficient: $D_x(E,t) =\frac{d}{dt}\Delta X^{ 2}(E_{F},t)$, which in the long time limit gives the conductivity through Eq.~\ref{eq:kubo}. Assuming an isotropic system in the $x$ and $y$ directions, the 2D diffusion coefficient becomes $D(t) = D_x(t) + D_y(t) = 2D_x(t)$. All information about multiple scattering effects is contained in the time-dependence of $D(t)$. 

Numerically, whatever the initial wavepacket features, $D(t)$ starts increasing ballistically at short times, then reaches a maximum value which depends on the disorder strength, and finally decays as a result of quantum interferences, the strength of which will dictate either a weak or a strong Anderson localization regime.  The semiclassical quantities (elastic mean free path $\ell_e(E)$ and semiclassical conductivity $\sigma_{sc}$) are derived from the maximum of $D(t)$ as $\ell_e(E)=D^{\text{max}}(E)/2 v(E)$ and $\sigma_{sc}(E)=\frac{1}{4}e^2 \rho(E) D^{\text{max}}(E)$, respectively (with $v(E)$ being the carrier velocity). 

The conductance of the amorphous stripes contacted to pristine graphene electrodes is computed using the Landauer-B\"uttiker approach \cite{buttiker}: 
\begin{equation}
G(E) = G_0 T(E) = {2e^2 \over h} {\large\rm Tr} \bigl[ t^\dagger t \bigr]
\end{equation}
where $T(E)$ and $t(E)$ are the transmission probability and transmission matrix, respectively, which can be computed from the Green's function $G(E)$ in the contact region and the broadening  $\Gamma(E)$ of the states due to the interaction with the left and right electrodes.
%\begin{equation} 
%t(E) = \Gamma_L (E)^{1/2} G(E) \Gamma_R (E)^{1/2}
%\end{equation}
We calculate the conductance of the stripes, which are infinite and periodic in the direction
parallel to the interface with the graphene electrodes (y axis in Fig.~\ref{dos}(c)).
Despite the very large periodicity of our models, we perform a thorough sampling of the $k_y$-points in that direction \cite{thygesen}, to obtain the appropriate V-shaped conductance of graphene in the thermodynamic limit. $G$ is given per supercell of periodicity $W$=11.4 nm. Note that, with this geometry, conductivity and conductance  are related though $\sigma={L \over W} G$.

\begin{figure}[htbp]
\begin{center}
\leavevmode
\includegraphics[scale=1.0]{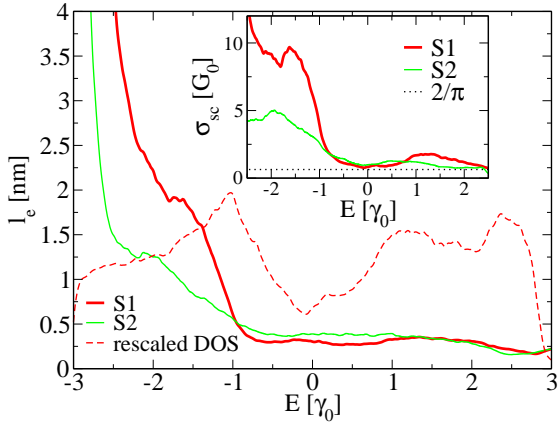}
\caption{(color online) Elastic mean free path versus energy for the two samples. DOS of sample S1 is also shown for comparison, in rescaled unts. Inset: semiclassical conductivity of corresponding lattices.}
\label{mfp}
\end{center}
\end{figure}

\vspace{1mm}
\textit{Mean Free Path, Conductivity and Localization Effects}.-
Fig. \ref{difconst} shows the time dependence of the normalized diffusion coefficient $D(t)/D_{\rm{max}}$  for two chosen energies and for both samples S1 and S2. For energy $E=-2\gamma_0$, the short-time ballistic regime is followed by the  saturation of the diffusion coefficient typically after $0.1$~ps (for both samples S1 and S2). From the saturation values, $\ell_e(E)$ and  $\sigma_{sc}$ are deduced and reported in Fig. \ref{mfp}. A striking feature is the very low value of the mean free path $\ell_{e}$, below $0.5$ nm for the energy window around the Fermi level (where the DOS is considerably larger than that of  pristine graphene). For negative energies (holes) far from the charge neutrality point, a considerable increase of more than one order of magnitude in the mean free paths is observed. The increase occurs for smaller binding energies for sample S1 than for sample S2, in good correlation with the changes observed in the DOS (which, around the van Hove singularity, deviates from the pristine graphene one more strongly for sample S2).  

We observe in the inset of Fig.~\ref{mfp} that $\sigma_{sc}$ shows a minimum value $\sigma_{sc}^{\rm min}$ about $4e^{2}/\pi h$, in agreement with the theoretical limit in the diffusive regime,  already confirmed for other types of disorder \cite{Minimcond,Minimcond2}. However, in contrast to prior studies,  conductivity values near the minimum are obtained over an energy range of several eV around the charge neutrality point. This indicates that transport is strongly degraded in the amorphous network compared to pristine graphene, for which the conductivity increases rapidly when the Fermi level is shifted away from the Dirac point.  The charge mobility, $\mu(E)=\sigma_{sc}(E)/en(E)$, with $n(E)$ being the carrier density, is found to be about $10$ {\rm cm}$^{2}${\rm V}$^{-1}${\rm s}$^{-1}$ for $n=10^{11}-10^{12}$~cm$^{-2}$, orders of magnitudes lower than those usually measured in graphene samples \cite{Mob}. Such low conductivity and mobility values should be measured at room temperature, where the semiclassical approximation is expected to hold.

The obtained short mean free paths and minimum (semiclassical) conductivities indicate a further marked contribution of quantum interferences turning the system to a weak and strong insulating system with temperature drop. Interference effects are evidenced by the time-dependent decay of the diffusion coefficient $D(t)/D_{\rm{max}}$ as clearly seen in Fig. \ref{difconst}, although with large differences depending on the chosen energy. For $E=-2\gamma_0$, such decay is weak but clearly more pronounced for the S2 sample (which is more disordered than S1). In sharp contrast, localization effects are much stronger at the charge neutrality point and develop from much shorter timescale (few nanoseconds). These differences will be reflected in the corresponding localization lengths.  

Based on the scaling theory of localization, an estimate of the localization length of electronic states is inferred from $\xi(E)=\ell_{e}(E)\exp(\pi h\sigma_{sc}(E)/2e^{2})$ \cite{lee}. The results are shown in Fig. \ref{difconst} (inset). The amorphous samples are confirmed to be extremely poor conductors, with localization lengths as low as $\xi\sim 5-10$ nm over a large energy window around the charge neutrality point. One also observes that $\xi$ can vary by more than one order of magnitude depending on the disordered topology of the sample and rings statistics.

\begin{figure}[htbp]
\begin{center}
\leavevmode
\includegraphics[scale=0.3]{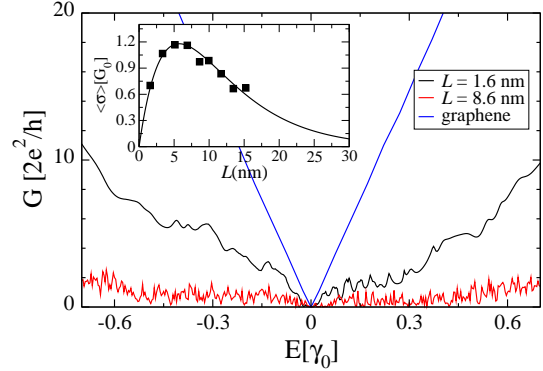}
\caption{(color online) Landauer-B\"uttiker conductance (for $W$=11.4 nm) of two amorphous  stripes contacted to graphene electrodes with $L = 1.6$ and 8.6 nm, respectively. The conductance of a pristine graphene contact with the same lateral size (11.4 nm) is shown for comparison. The inset shows the dependence of the conductivity on the  stripe size $L$; symbols: calculated points; line: fit to    $\sigma(L) \sim {L \over W} e^{-L/\xi}$.}
\label{landauer}
\end{center}
\end{figure}

To further confirm the localization lengths estimated using scaling theory, we compute explicitly the conductance of the amorphous graphene  stripes contacted with pristine graphene electrodes, using the geometry shown in Fig.~\ref{dos}(c), as a function of the length of the amorphous contact $L$. Fig. \ref{landauer} shows the conductance curves for two  stripes with $L=1.6$ and 8.6 nm, respectively, compared to that of a graphene contact with the same lateral size in the supercell ($W=11.4$~nm). 
%Note that the use of the $\Gamma$ point   to sample $k_\parallel$ reflects in the stepwise conductance of pristine graphene. 
It is clear that the conductance of the amorphous samples is greatly reduced with respect to that of graphene, and that the reduction is more pronounced as the length of the amorphous contact increases. Also, while the conductance for the  stripe with the smallest length is relatively smooth, it becomes more noisy with increasing $L$. This reflects the transition from a diffusive system, in which the length of the amorphous contact is longer than the mean free path but shorter than the localization length, to a strongly localized one in which the localization length is shorter than the length of the amorphous region.

From the variation of the Landauer-B\"uttiker conductance with $L$, we can extract reliable values of the localization lengths, as in the Anderson regime the conductance should decay as $G(L) \sim e^{-L/\xi}$. The inset in Fig. \ref{landauer} shows the value of the conductivity (obtained from the Landauer-B\"utikker conductance) for several stripes with $L$ ranging from 1.6 to 15.3 nm, averaged over an energy window of 1.5$\gamma_0$ around the Fermi energy.  A fit of the results to $\sigma(L) \sim {L \over W} e^{-L/\xi}$ yields  $\xi=5.8$ nm. This value is fully consistent with that obtained above using scaling theory for energies close to the Fermi level, and confirms that, in these amorphous structures, strong localization effects should occur at low temperatures at distances of less than 10 nm.  Our results are consistent with the experimental ones from transport measurements by Zhou {\em et al.} \cite{zhou}, which show values in the range between 0.1 and 10 nm for samples amorphized by ion radiation.

In conclusion, we have shown that topological disorder alone causes amorphous graphene to be a strong Anderson insulator. The increase of the density of states close to the charge neutrality point is associated with quantum interference which inhibits current flow at low temperatures.  Very short mean free paths and localization lengths are predicted, in line with recent experimental evidence in graphene under heavy ion irradiation damage \cite{zhou}.

P.O. acknowledges support from Spanish MICINN (Grants FIS2009-12721-C04-01 and CSD2007-00050), M.F.T from the US National Science Foundation under grant DMR-0703973, and A.K. from the US Department of Education under grant GAANN P200A090123 and the ARCS Foundation. This work is supported by the European Community through the Marie Curie Actions.

\end{document}